\documentclass[aps,prd,amsmath,one column,notitlepage,showpacs,superscriptaddress,nofootinbib,usenatbib,11pt]{revtex4-1}
\def \be {\begin{equation}} 
\def \ee {\end{equation}} 
\def \bea {\begin{eqnarray}} 
\def \eea {\end{eqnarray}} 

\usepackage{graphicx}
\usepackage{dcolumn}
\usepackage{bm}
\usepackage{epsfig} 
\usepackage{amsfonts}
\usepackage{amsmath}
\usepackage{amssymb}
\usepackage[usenames]{color}
\usepackage[dvipsnames]{xcolor}
\usepackage[unicode, colorlinks=true, linkcolor=linkcolor, citecolor=linkcolor, filecolor=linkcolor,urlcolor=linkcolor, pdfusetitle]{hyperref}
\usepackage{listings}
\usepackage[T1]{fontenc}

\hypersetup{colorlinks,citecolor=blue,linkcolor=blue,urlcolor=blue}
\hypersetup{final=true}

\definecolor{lgreen}{HTML}{079d90}

\begin{document}

\title{A model-independent test of speed of light variability with cosmological observations}

\author{Gabriel Rodrigues}
\email{gabrielrodrigues@on.br}
\affiliation{Observat\'orio Nacional, 20921-400, Rio de Janeiro - RJ, Brazil}

\author{Carlos Bengaly}
\email{carlosbengaly@on.br}
\affiliation{Observat\'orio Nacional, 20921-400, Rio de Janeiro - RJ, Brazil}

\date{\today}

\begin{abstract}
A powerful test of fundamental physics consists on probing the variability of fundamental constants in Nature. Although they have been measured on Earth laboratories and in our Solar neighbourhood with extremely high precision, it is crucial to carry out these tests at the distant Universe, as any significant variation of these quantities would immediately hint at new physics. We perform a cosmological measurement of the speed of light using the latest Type Ia Supernova and cosmic chronometer observations at the redshift range $0<z<2$. Our method relies on the numerical reconstruction of these data in order to circumvent {\it a priori} assumptions of the underlying cosmology. We confirm the constancy of the speed of light at such redshift range, reporting two $\sim 5$\% precision measurements of $c = (3.20 \pm 0.16) \; \times 10^5 \; \mathrm{km \; s}^{-1}$ in $z \simeq 1.58$, and $c = (2.67 \pm 0.14) \; \times 10^5 \; \mathrm{km \; s}^{-1}$ in $z \simeq 1.36$, depending on the reconstruction method, at a $1\sigma$ confidence level.
\end{abstract}

\maketitle



\section{Introduction}\label{sec:intro}

The standard cosmological model (SCM) consists on the flat $\Lambda$CDM paradigm since the late 1990s~\cite{riess98, perlmutter99}. This model describes a Universe dominated by cold dark matter, which is responsible for structure formation and galaxy dynamics, and the Cosmological Constant $\Lambda$ - an exotic fluid that is responsible for the late-time cosmic acceleration. Such scenario is based on two fundamental pillars: (a) the assumption of a Friedmann-Lema\^itre-Robertson-Walker (FLRW) Universe at large scales, ~\cite{wu99, clarkson10, maartens11, clarkson12}, and (b) General Relativity (GR) as the underlying theory of gravity~\cite{clifton12, will14, berti15, coley20}. Recent observations of Cosmic Microwave Background (CMB)~\cite{planck21}, Type Ia Supernovae (SNe) luminosity distances~\cite{pantheon18}, besides the clustering and weak lensing of cosmic structures~\cite{eboss21, kids21, des21a, des21b} confirmed this scenario with unprecedented precision. Nonetheless, this model suffers from unsolved issues - such as the fine-tuning and the primordial singularity problems, to name a few. As of recently, tensions between cosmological parameter measurements within its framework have been persisting - the most stringent at the moment being the $(4.5-5.0)\sigma$ tension between the $H_0$ measured in the late- and early-time Universe from SNe and CMB, respectively~\cite{divalentino21, shah21, riess21}.

It is therefore crucial to propose and test alternative models and revisit the foundations of the SCM. Any statistically significant departure from its predictions would hint at new physics and lead to a complete reformulation of its scenario. Recent works confirmed the validity of both FLRW and GR at large cosmological scales, yet some results may state otherwise - see ~\cite{perivolaropoulos21} for a recent review on unsolved cosmological puzzles. 

Another possible route consists on probing the variability of fundamental constants of Nature as a test of fundamental physics~\cite{dirac37, uzan03, uzan11, martins17}. Experiments on Earth and Solar System have been designed and carried out for centuries to measure these quantities, with null results for their variations - and extreme precision in their measurements. Still, cosmological tests of fundamental constants variations are much more scarce and less precise, given the difficulty in obtaining cosmological data in high redshifts that would allow the conduction of such tests. 

Care should be taken when developing such models, as they may lead to further issues in physical laws that were constructed on these physical constants. Varying speed of light $c$ models (VSL), for instance, must be able to reproduce the success of special relativity in explaining electromagnetism and thermodynamics, at least. Some of them satisfy these requirements and may provide viable solutions to the SCM issues, e.g. the horizon and flatness problem at primordial times, as well as the cosmic coincidence and the $H_0$ tension in the late-time Universe~\cite{moffat93, barrow98, albrecht99, barrow99a, barrow99b, clayton99, avelino99, clayton00, bassett00, magueijo00, clayton02, magueijo03, ellis05, ellis07, magueijo08, cruz12, moffat16, franzmann17, cruz18, costa19, gupta20, lee21a, lee21b, lee21c, lee21d, lee21e}. 

Under this motivation, we perform a test of the speed of light variance with the latest cosmological observations. Some recent analyses focused on this topic reported no significant evidence for VSL~\cite{balcerzak14a, zhang14, balcerzak14b, qi14, salzano15, salzano16, cai16, balcerzak17, cao17, salzano17a, salzano17b, guedeslang18, zou18, martinez-huerta18, cao18, guedeslang19, wang19, HAWC20, pan20, nguyen20, gupta21, agrawal21, mendonca21}. Due to the caveats of the SCM just discussed, our goal is to revisit these tests using different method and data-sets in order to search for new physics that would help solving (or ruling out) this framework. 

The paper is organised as follows: We describe the method and data deployed in our analysis in Section 2, section 3 presents our results, and section 4 is dedicated to the discussion and concluding remarks.  


\section{Method}\label{sec:data_meth}


\subsection{Measuring the speed of light}

We adopt the method proposed by~\cite{salzano15} to measure $c$ and test the evidence for VSL with cosmological observations. Assuming a flat FLRW Universe, the angular diameter distance is given by~\cite{hogg99}
\begin{equation}\label{eq:DAz}
    D_{\rm A}(z)=\frac{1}{(1+z)}\int_0^z\frac{cdz}{H(z)} \,,
\end{equation}
where $c=c(z)$ is the speed of light as function of redshift. 
Differentiating Eq.~\eqref{eq:DAz} w.r.t to $z$, we obtain
\begin{equation}\label{eq:diff_DAz}
    \frac{\partial}{\partial z}[(1+z)D_{\rm A}(z)]=\frac{c(z)}{H(z)} \,,
\end{equation}
so that $c(z)$ reads
\begin{equation}\label{eq:cz}
    c(z) = H(z)[(1+z)D'_{\rm A}(z) + D_{\rm A}(z)] \,,
\end{equation}
where the prime denotes a first order derivative in redshift. Its respective uncertainty can be obtained through standard error propagation, as follows
\begin{eqnarray}\label{eq:errcz}
       \sigma^2_{c(z)} = {H'(z)[(1+z)D'_{\rm A}(z) + D_{\rm A}(z)]\sigma_{H(z)}}^2 + {H(z)[(1+z)D''_{\rm A}(z) + D_{\rm A}(z)]\sigma_{D'_{\rm A}(z)}^2} \nonumber \\ + {H(z)[(2+z)D'_{\rm A}(z)]\sigma_{D_{\rm A}(z)}^2}
\end{eqnarray}
It is well known that the angular diameter distance reaches a maximum value at some given redshift~\cite{weinberg72}~\footnote{This redshift value would be $z=1.592^{+0.043}_{-0.039}$ for a flat $\Lambda$CDM model assuming Planck latest constraints~\cite{salzano15}.}. Hence, the first order derivative of the angular diameter distance is going to be zero, so from Eq.~\eqref{eq:cz} we get that
\begin{equation}\label{eq:czm}
    c(z_{\rm m})=D_{\rm A}(z_{\rm m})H(z_{\rm m}) \,,
\end{equation}
where $z_{\rm m}$ corresponds to the redshift value where it occurs.

We note that this approach is only valid for flat FLRW models, since there is a degeneracy between a VSL and the cosmic curvature~\cite{salzano16, salzano17a}. Current constraints on the latter are in excellent agreement with a null curvature value~\cite{planck21, pantheon18, eboss21, kids21, des21a, des21b} (yet debated by~\cite{divalentino19,handley21}), thus we can safely make this assumption. A more thorough examination of this degeneracy is left for future works.


\subsection{Data analysis}

Our analysis uses 1048 distance measurements from the Pantheon SN compilation~\cite{pantheon18}\footnote{We note that the SN light-curves depend on the assumption of constant speed of light, which is encompassed in the Chandrasekhar limit.}, in addition to 50 Hubble parameter measurements $H(z)$ obtained from galaxy age and the radial mode of baryonic acoustic oscillations (BAO), as compiled by~\cite{magana18, moresco22}~\footnote{We note that the covariances of radial BAO and SN distance measurements were neglected in our analysis for the sake of computational issues.}. Although transverse BAO measurements would be a more natural choice for such analysis, since it consists on a statistical standard ruler, there are very few data points at $z \sim 1.5$. Hence we use 1048 SN apparent magnitude measurements $m_{\rm B}$ combined with the absolute magnitude determination~\cite{pantheon18}
\begin{equation}\label{eq:Mb}
M_{\rm B}=-19.248 \pm 0.029 \,,  
\end{equation}
in order to obtain the luminosity distance $D_{\rm L}(z)$, given by
\begin{equation}\label{eq:DL}
D_{\rm L} = 10^{\frac{m_{\rm B} - M_{\rm B} - 25}{5}} \,,  \end{equation}
which can be converted into $D_{\rm A}(z)$ through the cosmic distance duality relation (CDDR), which follows
\begin{equation}\label{eq:cddr}
   D_{\rm A}(z)=D_{\rm L}(z)(1+z)^{-2} \,.
\end{equation}
Probing CDDR consists on one of the most crucial tests of fundamental Cosmology, since it {\emph does not} depend on dark energy models, GR, or even the FLRW hypothesis. The CDDR can only be violated in case of non-Riemmanian geometry, the presence of a cosmic opacity source, or variations of fundamental physics e.g. the fine-structure constant and the Equivalence Principle, to name a few~\cite{lee21d}. Because severe constraints have been imposed on departures of the CDDR using a plethora of approaches and data-sets~\cite{bassett04, holanda10, holanda12, santos-da-costa15, holanda16, goncalves20, mukherjee21, bora21, holanda21}, we can reliably assume the validity of Eq.~\eqref{eq:cddr} to convert $D_{\rm L}(z)$ measurements into $D_{\rm A}(z)$. 

We reconstruct the $D_{\rm A}(z)$ and $H(z)$ data points using a non-parametric approach based on the Gaussian Processes method with the {\sc GaPP} package~\cite{seikel12} - see also~\cite{shafieloo12}. Similar approaches have been extensively explored in the literature for the sake of obtaining cosmological constraints without making {\it a priori} assumptions on the material content of the Universe~\cite{yahya13, sapone14, busti14, gonzalez16, marra18, gomezvalent18a, gomezvalent18b, bengaly21a, benisty21, mukherjee20a, mukherjee20b, bonilla21, vonMarttens21, colgain21, arjona21, perenon21, bernardo21a, bernardo21b, bengaly21b, bernardo21c, dialektopoulos21, avila22}. Therefore, such a method provide robust evidence for the reconstructed quantities of interest, making GP suitable for tests like the one we pursue here. We adopt the squared exponential (hereafter SqExp) and Mat\'ern(9/2) (hereafter Mat92) kernels as our default choices throughout this analysis, so we can make a direction comparison between two different reconstruction methods. We assume $n=1000$ reconstruction bins along the $0<z<2$ range to get the $D_{\rm A}(z)$, $H(z)$, and their respective derivatives reconstructions. Also, we opt not to optimise the GP hyperparameters in order to avoid possible biases in these reconstructions due to overfitting.

\section{Results}

\begin{figure*}[!t]
\centering
\includegraphics[width=0.49\textwidth, height=6.0cm]{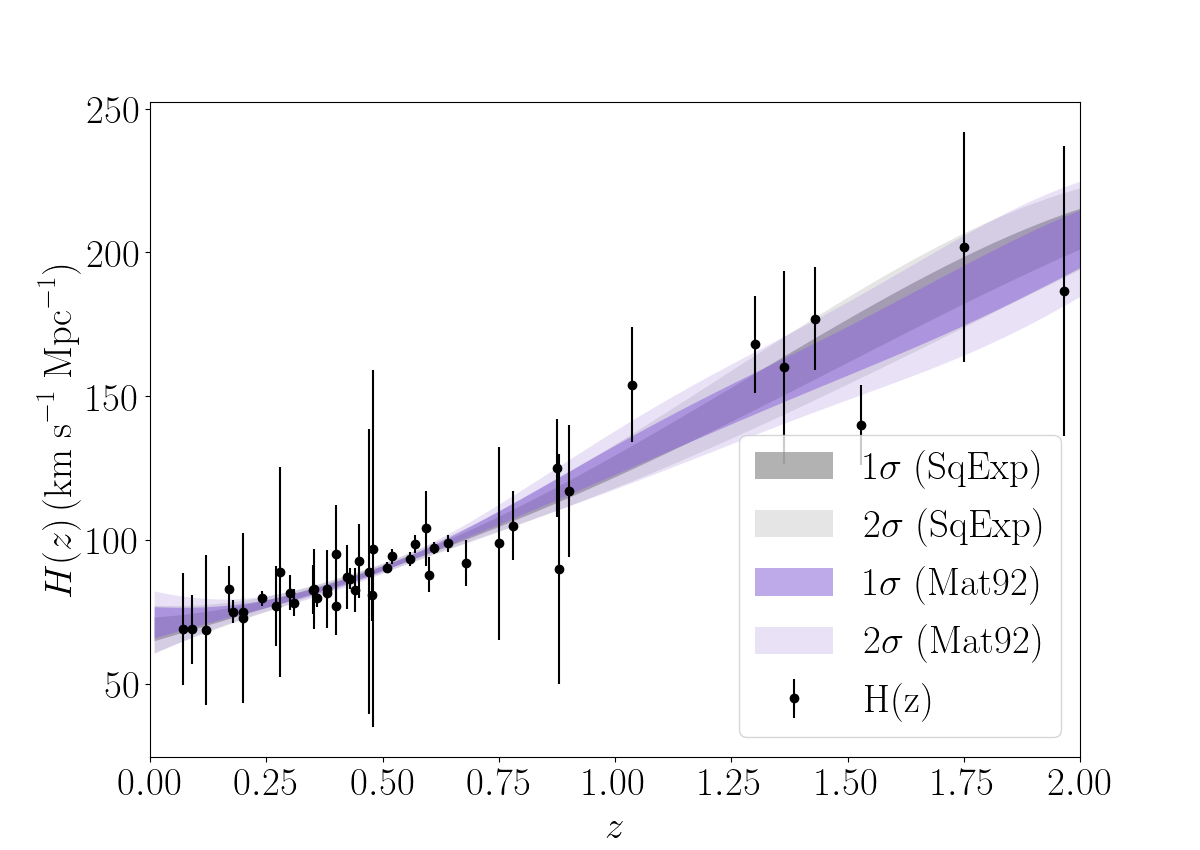}  
\includegraphics[width=0.49\textwidth, height=6.0cm]{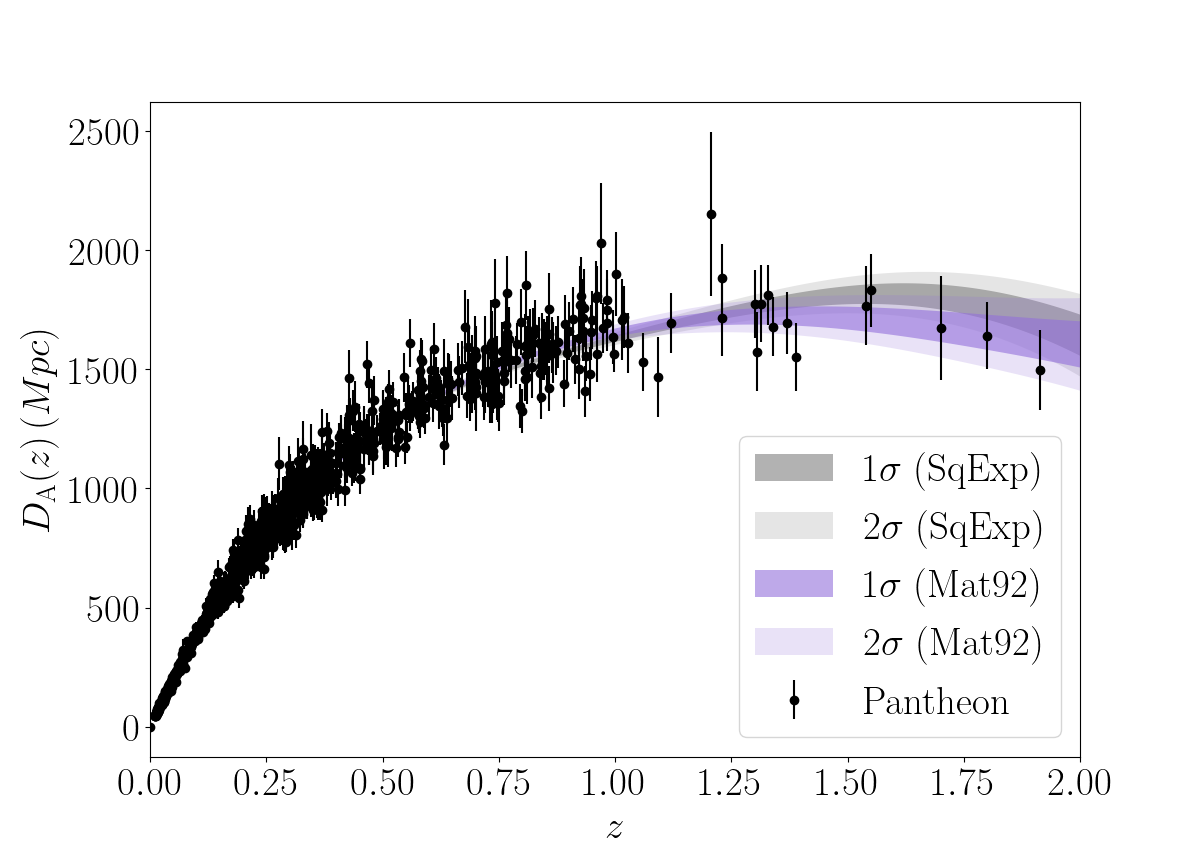} 
\includegraphics[width=0.49\textwidth, height=6.0cm]{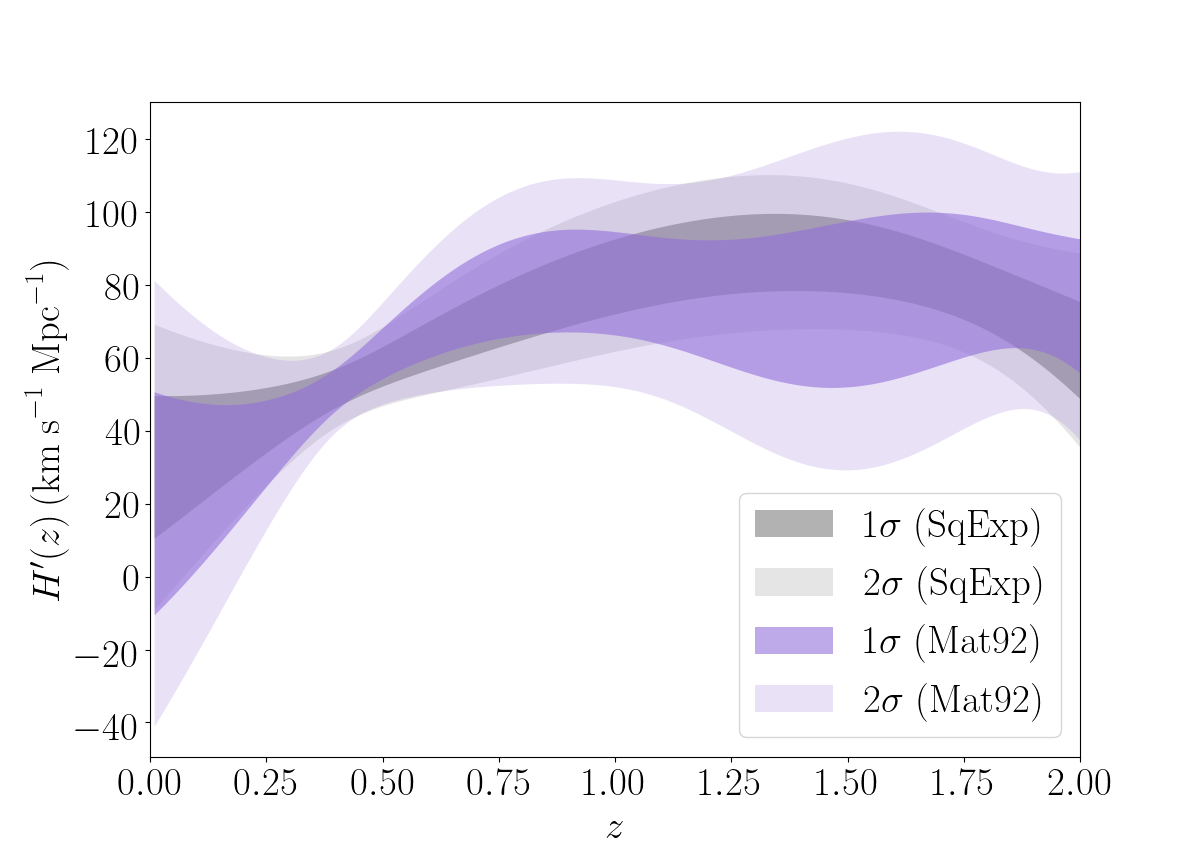}  
\includegraphics[width=0.49\textwidth, height=6.0cm]{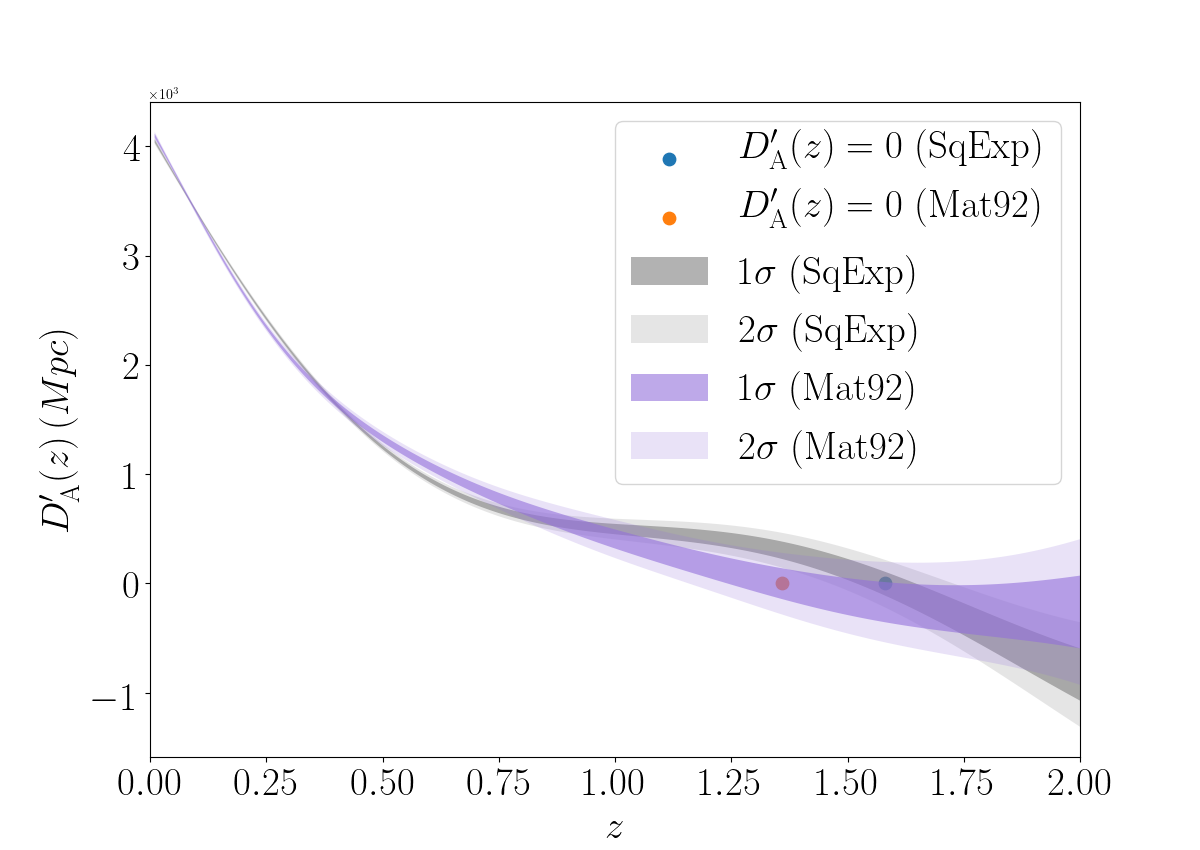}
\caption{The upper left panel shows the 50 $H(z)$ data points along with their respective GP reconstruction using the SqExp (grey shade) and Mat92 kernel (purple shade). The upper right panel displays the angular diameter distance $D_\mathrm{A}(z)$ points from SNe obtained through the CDDR, as well as their respective GP reconstructed curves. The lower left panel present the first order derivative of the Hubble parameter, namely $H'(z)$, whereas the $D'_\mathrm{A}(z)$ is shown in the lower right panel. The blue and orange dots denote the redshift where $D'_\mathrm{A}(z)=0$ for the SqExp and Mat92 kernels, respectively. The darker (lighter) shaded areas provide the $1\sigma$ ($2\sigma$) confidence regions.}
\label{fig:hz_daz}
\end{figure*}

\begin{figure*}[!t]
\includegraphics[width=0.98\textwidth]{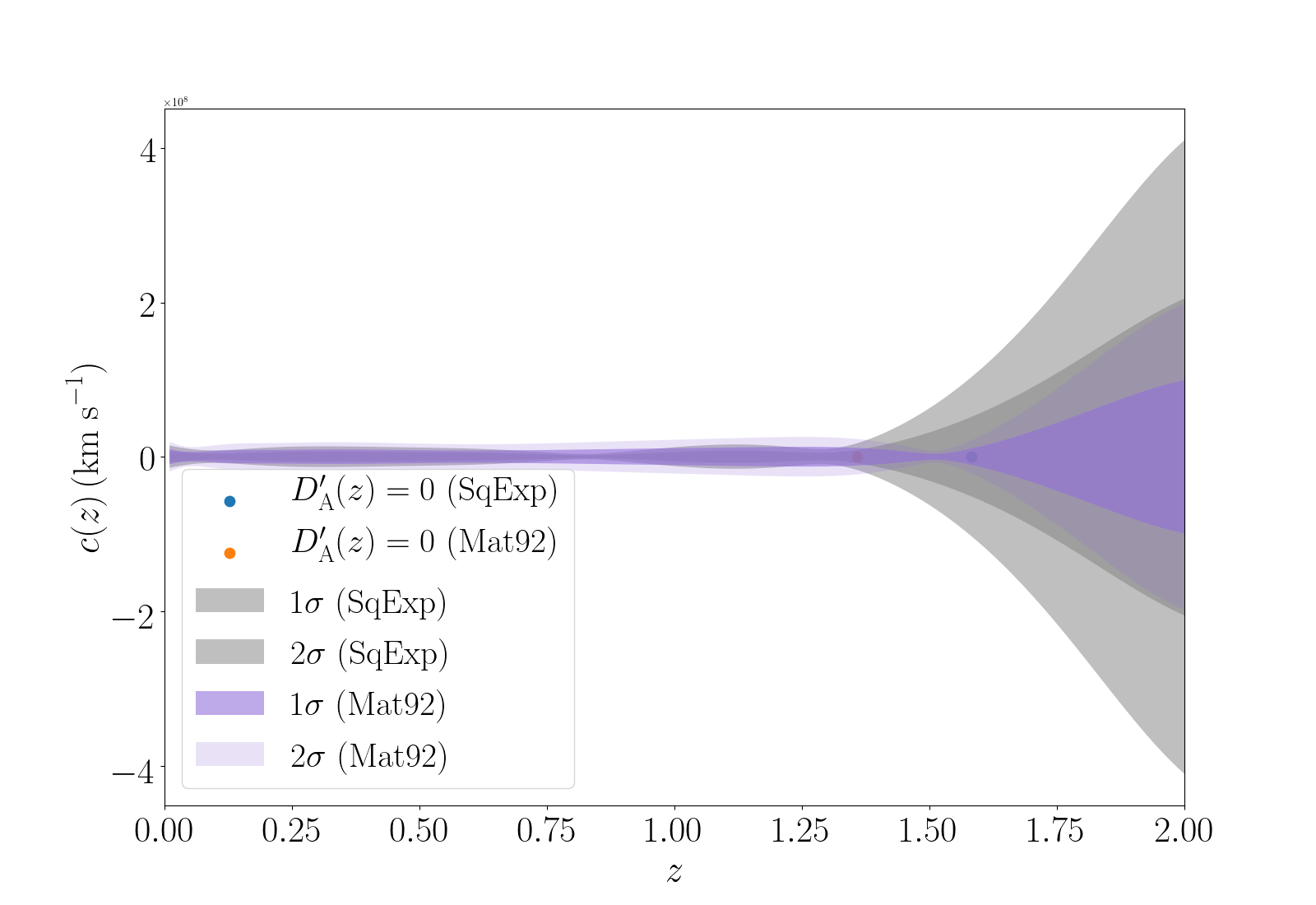} 
\caption{The speed of light $c(z)$ reconstruction across the redshift range $0<z<2$ for both SqExp (grey) and Mat92 (purple) kernels. Again, the shaded areas correspond to the $1\sigma$ and $2\sigma$ confidence regions.}
\label{fig:cz}
\end{figure*}

We show the $1$ and $2\sigma$ reconstructed curves of $H(z)$, $D_{\rm A}(z)$, and their respective first order derivatives in Fig.~\ref{fig:hz_daz}. The upper left panel displays the 50 Hubble parameter $H(z)$ measurements from galaxy age plus radial BAO data, whereas the upper right panel presents the 1048 $D_{\rm A}(z)$ measurements from the Pantheon SN compilation under the CDDR assumption. Both plots show their respective GP reconstructions as well, where the grey shaded areas denote those obtained using the SqExp kernel, and the purple ones correspond to the Mat92. Darker and lighter shades correspond to $1$ and $2\sigma$ uncertainties of these reconstructions, respectively. In addition, we exhibit the $H'(z)$ and $D'_{\rm A}(z)$ reconstructions in the lower panels of the same figure. The blue (orange) dot in the latter plot refers to the redshift where the angular diameter distance reaches its maximum value, i.e., $D'_{\rm A}(z_{\rm m})=0$, for both kernel choices. These values are $z_{\rm m} \simeq 1.58$ (SqExp) and $z_{\rm m} \simeq 1.36$ (Mat92).

In Fig.~\ref{fig:cz}, we show the speed of light evolution w.r.t. redshift obtained from Eqs.~\eqref{eq:cz} and~\eqref{eq:errcz}. Again, the grey and purple shades denote the SqExp and Mat92 reconstructions, and the blue and orange dots correspond to $z_{\rm m}$, i.e., the point where we measure the speed of light according to equation~\ref{eq:czm}, assuming the SqExp and Mat92 kernels, respectively. We obtain $c(1.58) = (3.20 \pm 0.16) \; \times 10^5 \; \mathrm{km \; s}^{-1}$, and $c(1.36) = (2.67 \pm 0.14) \; \times 10^5 \; \mathrm{km \; s}^{-1}$ at a $1\sigma$ confidence level, which amount to 5.0\% and 5.3\% precision measurements, respectively. When $z>z_{\rm m}$, our reconstruction rapidly degrades due to insufficient data, as we can see in the lower panels of Fig.~\ref{fig:hz_daz}.

Our results agree with local measurements of the speed of light at a $\sim 1.5\sigma$ confidence level, and also with previous works, such as $c=(3.04 \pm 0.18) \; \times 10^5 \; \mathrm{km \; s}^{-1}$ at $1\sigma$ reported by~\cite{cao17} using angular diameter distance measurements from quasars. We found no evidence for VSL models as well at the $0<z<2$ range, as evident in Fig.~\ref{fig:cz}. Furthermore, we note that these results are robust with respect to other GP kernel, such as the double squared exponential and Mat\'ern(7/2) - which gave almost identical results as the SqExp and Mat92 kernels, respectively - as well as different binning schemes, like $n=100$ or $n=2000$.


\section{Concluding remarks}

Testing the variability of fundamental constants in Nature consists in one of the strongest tests of fundamental Physics. Any significant evolution of these values would immediately hint at new physics, and demand a profound reformulation of the standard model of Cosmology and Particles, not to mention Electromagnetism, Thermodynamics and Gravitation. Such tests have been thoroughly carried out on Earth and its surrounding Solar neighbourhood for decades with exquisite precision, providing no evidence for their evolution. However, cosmological tests of this kind are still sparse given the difficulty of taking precise observations at high redshifts. Hence, there is an urge to seek them as a means to challenge the validity of the standard cosmological model. 

That serves us as a motivation to test the variance of the speed of light using cosmological data of cosmic chronometers and standardisable candles. We deploy a model-independent analysis using Gaussian Processes to reconstruct the Hubble parameter from galaxy age and radial BAO mode, and angular diameters distances from Type Ia Supernovae under the assumption of the cosmic distance duality relation. We follow the approach proposed by Salzano et al.~\cite{salzano15}, but relying on different data-sets and approaches than formerly pursued in the literature. 

We obtained a $\sim$5\% precision measurements of the speed of light, i.e., $c = (3.20 \pm 0.16) \; \times 10^5 \; \mathrm{km \; s}^{-1}$ at $z_{\rm m} \simeq 1.58$, assuming the squared exponential Gaussian Process kernel, whilst the Mat\'ern(9/2) kernel gave $c = (2.67 \pm 0.14) \; \times 10^5 \; \mathrm{km \; s}^{-1}$ at $z_{\rm m} \simeq 1.36$. Both results indicate no evidence for speed of light variability at the redshift range $0<z<2$, which covers a light-travel time of nearly $10.5$ Gyrs. We also note that the $c(z)$ reconstruction significantly degrades at $z>z_{\rm m}$ due to the lack of observational data at such redshift range - hence stressing the necessity of further development in cosmological measurements to improve such measurements. We plan to repeat this analysis with simulated measurements of forthcoming surveys such as J-PAS~\cite{benitez14, aparicioresco20} in a future work.

Overall, our findings are fully consistent with the fundamental assumption of null variance of fundamental physical constants, therefore helping to underpin the foundations of the standard cosmological model~\cite{peebles21}. 

\newpage

\emph{Acknowledgments} -- 
GR and CB acknowledge Rodrigo Von Marttens, Javier Gonzalez and Jailson Alcaniz for insightful discussions. GR acknowleges financial support by CAPES. CB was supported by the Programa de Capacita\c{c}\~ao Institucional PCI/ON fellowship at an early stage of this work, and FAPERJ postdoc nota 10 fellowship



\begin{thebibliography}{99}
\frenchspacing

\bibitem{riess98}
    A.~G.~Riess \textit{et al.} [Supernova Search Team],
    ``Observational evidence from supernovae for an accelerating universe and a cosmological constant,''
    Astron. J. \textbf{116} (1998), 1009-1038
    [arXiv:astro-ph/9805201].

\bibitem{perlmutter99}
    S.~Perlmutter \textit{et al.} [Supernova Cosmology Project],
    ``Measurements of $\Omega$ and $\Lambda$ from 42 high redshift supernovae,''
    Astrophys. J. \textbf{517} (1999), 565-586
    [arXiv:astro-ph/9812133].

\bibitem{wu99}
    K.~K.~S.~Wu, O.~Lahav and M.~J.~Rees,
    ``The large-scale smoothness of the Universe,''
    Nature \textbf{397} (1999), 225-230
    [arXiv:astro-ph/9804062].

\bibitem{clarkson10}
    C.~Clarkson and R.~Maartens,
    ``Inhomogeneity and the foundations of concordance cosmology,''
    Class. Quant. Grav. \textbf{27} (2010), 124008
    [arXiv:1005.2165].

\bibitem{maartens11}
    R.~Maartens,
    ``Is the Universe homogeneous?,''
    Phil. Trans. Roy. Soc. Lond. A \textbf{369} (2011), 5115-5137
    [arXiv:1104.1300].

\bibitem{clarkson12}
    C.~Clarkson,
    ``Establishing homogeneity of the universe in the shadow of dark energy,''
    Comptes Rendus Physique \textbf{13} (2012), 682-718
    [arXiv:1204.5505].
    
\bibitem{clifton12}
    T.~Clifton, P.~G.~Ferreira, A.~Padilla and C.~Skordis,
    ``Modified Gravity and Cosmology,''
    Phys. Rept. \textbf{513} (2012), 1-189
    [arXiv:1106.2476].
    
\bibitem{will14}
    C.~M.~Will,
    ``The Confrontation between General Relativity and Experiment,''
    Living Rev. Rel. \textbf{17} (2014), 4
    [arXiv:1403.7377].

\bibitem{berti15}
    E.~Berti, E.~Barausse, V.~Cardoso, L.~Gualtieri, P.~Pani, U.~Sperhake, L.~C.~Stein, N.~Wex, K.~Yagi and T.~Baker, \textit{et al.}
    ``Testing General Relativity with Present and Future Astrophysical Observations,''
    Class. Quant. Grav. \textbf{32} (2015), 243001
    [arXiv:1501.07274].

\bibitem{coley20}
    A.~A.~Coley and G.~F.~R.~Ellis,
    ``Theoretical Cosmology,''
    Class. Quant. Grav. \textbf{37} (2020) no.1, 013001
    [arXiv:1909.05346].
    
\bibitem{planck21} 
  N.~Aghanim {\it et al.} [Planck Collaboration],
   ``Planck 2018 results. VI. Cosmological parameters,''
   Astron. Astrophys. \textbf{641} (2020), A6
  [erratum: Astron. Astrophys. \textbf{652} (2021), C4]
  [arXiv:1807.06209].
    
\bibitem{pantheon18} 
  D.~M.~Scolnic {\it et al.},
  ``The Complete Light-curve Sample of Spectroscopically Confirmed SNe Ia from Pan-STARRS1 and Cosmological Constraints from the Combined Pantheon Sample,''
  Astrophys.\ J.\  {\bf 859}, no. 2, 101 (2018)
  [arXiv:1710.00845].

\bibitem{eboss21}
    S.~Alam \textit{et al.} [eBOSS],
    ``Completed SDSS-IV extended Baryon Oscillation Spectroscopic Survey: Cosmological implications from two decades of spectroscopic surveys at the Apache Point Observatory,''
    Phys. Rev. D \textbf{103} (2021) no.8, 083533
    [arXiv:2007.08991].

\bibitem{kids21}
    C.~Heymans, T.~Tr\"oster, M.~Asgari, C.~Blake, H.~Hildebrandt, B.~Joachimi, K.~Kuijken, C.~A.~Lin, A.~G.~S\'anchez and J.~L.~van den Busch, \textit{et al.}
    ``KiDS-1000 Cosmology: Multi-probe weak gravitational lensing and spectroscopic galaxy clustering constraints,''
    Astron. Astrophys. \textbf{646} (2021), A140
    [arXiv:2007.15632].
    
\bibitem{des21a}
    T.~M.~C.~Abbott \textit{et al.} [DES],
    ``Dark Energy Survey Year 3 Results: Cosmological Constraints from Galaxy Clustering and Weak Lensing,''
    Phys. Rev. D \textbf{105} (2022) no.2, 023520
    [arXiv:2105.13549].

\bibitem{des21b}
    L.~F.~Secco \textit{et al.} [DES],
    ``Dark Energy Survey Year 3 Results: Cosmology from Cosmic Shear and Robustness to Modeling Uncertainty,''
    Phys. Rev. D \textbf{105} (2022) no.2, 023515[arXiv:2105.13544].
    
\bibitem{divalentino21}
    E.~Di Valentino, O.~Mena, S.~Pan, L.~Visinelli, W.~Yang, A.~Melchiorri, D.~F.~Mota, A.~G.~Riess and J.~Silk,
    ``In the realm of the Hubble tension\textemdash{}a review of solutions,''
    Class. Quant. Grav. \textbf{38} (2021) no.15, 153001
    [arXiv:2103.01183].

\bibitem{shah21}
    P.~Shah, P.~Lemos and O.~Lahav,
    ``A buyer's guide to the Hubble Constant,''
    Astron. Astrophys. Rev. \textbf{29} (2021) no.1, 9
    [arXiv:2109.01161].
    
\bibitem{riess21}
    A.~G.~Riess, W.~Yuan, L.~M.~Macri, D.~Scolnic, D.~Brout, S.~Casertano, D.~O.~Jones, Y.~Murakami, L.~Breuval and T.~G.~Brink, \textit{et al.}
    ``A Comprehensive Measurement of the Local Value of the Hubble Constant with 1 km/s/Mpc Uncertainty from the Hubble Space Telescope and the SH0ES Team,''
    [arXiv:2112.04510].

\bibitem{perivolaropoulos21}
    L.~Perivolaropoulos and F.~Skara,
    ``Challenges for $\Lambda$CDM: An update,''
    [arXiv:2105.05208].
    
\bibitem{dirac37}
    P.~A.~M.~Dirac,
    ``The Cosmological constants,''
    Nature \textbf{139} (1937), 323

\bibitem{uzan03}
    J.~P.~Uzan,
    ``The Fundamental Constants and Their Variation: Observational Status and Theoretical Motivations,''
    Rev. Mod. Phys. \textbf{75} (2003), 403
    [arXiv:hep-ph/0205340].
    
\bibitem{uzan11}
    J.~P.~Uzan,
    ``Varying Constants, Gravitation and Cosmology,''
    Living Rev. Rel. \textbf{14} (2011), 2
    [arXiv:1009.5514].
    
\bibitem{martins17}
    C.~J.~A.~P.~Martins,
    ``The status of varying constants: a review of the physics, searches and implications,''
    [arXiv:1709.02923].
    
\bibitem{moffat93}
    J.~W.~Moffat,
    ``Superluminary universe: A Possible solution to the initial value problem in cosmology,''
    Int. J. Mod. Phys. D \textbf{2} (1993), 351-366
    [arXiv:gr-qc/9211020].
    
\bibitem{barrow98}
    J.~D.~Barrow,
    ``Cosmologies with varying light speed,''
    [arXiv:astro-ph/9811022].

\bibitem{albrecht99}
    A.~Albrecht and J.~Magueijo,
    Phys. Rev. D \textbf{59} (1999), 043516
    [arXiv:astro-ph/9811018].

\bibitem{barrow99a}
    J.~D.~Barrow and J.~Magueijo,
    ``Solutions to the quasi-flatness and quasilambda problems,''
    Phys. Lett. B \textbf{447} (1999), 246
    [arXiv:astro-ph/9811073].

\bibitem{clayton99}
    M.~A.~Clayton and J.~W.~Moffat,
    ``Dynamical mechanism for varying light velocity as a solution to cosmological problems,''
    Phys. Lett. B \textbf{460} (1999), 263-270
    [arXiv:astro-ph/9812481].

\bibitem{barrow99b}
    J.~D.~Barrow and J.~Magueijo,
    ``Solving the flatness and quasiflatness problems in Brans-Dicke cosmologies with a varying light speed,''
    Class. Quant. Grav. \textbf{16} (1999), 1435-1454
    [arXiv:astro-ph/9901049].

\bibitem{avelino99}
    P.~P.~Avelino and C.~J.~A.~P.~Martins,
    ``Does a varying speed of light solve the cosmological problems?,''
    Phys. Lett. B \textbf{459} (1999), 468-472
    [arXiv:astro-ph/9906117].

\bibitem{clayton00}
    M.~A.~Clayton and J.~W.~Moffat,
    ``Scalar tensor gravity theory for dynamical light velocity,''
    Phys. Lett. B \textbf{477} (2000), 269-275
    doi:10.1016/S0370-2693(00)00192-1
    [arXiv:gr-qc/9910112].

\bibitem{bassett00}
    B.~A.~Bassett, S.~Liberati, C.~Molina-Paris and M.~Visser,
    ``Geometrodynamics of variable speed of light cosmologies,''
    Phys. Rev. D \textbf{62} (2000), 103518
    [arXiv:astro-ph/0001441].

\bibitem{magueijo00}
    J.~Magueijo,
    ``Covariant and locally Lorentz invariant varying speed of light theories,''
    Phys. Rev. D \textbf{62} (2000), 103521
    [arXiv:gr-qc/0007036].

\bibitem{clayton02}
    M.~A.~Clayton and J.~W.~Moffat,
    ``Vector field mediated models of dynamical light velocity,''
    Int. J. Mod. Phys. D \textbf{11} (2002), 187-206
    [arXiv:gr-qc/0003070].

\bibitem{magueijo03}
    J.~Magueijo,
    ``New varying speed of light theories,''
    Rept. Prog. Phys. \textbf{66} (2003), 2025
    [arXiv:astro-ph/0305457].

\bibitem{ellis05}
    G.~F.~R.~Ellis and J.~P.~Uzan,
    ```c' is the speed of light, isn't it?,''
    Am. J. Phys. \textbf{73} (2005), 240-247
    doi:10.1119/1.1819929
    [arXiv:gr-qc/0305099].
    
\bibitem{ellis07}
    G.~F.~R.~Ellis,
    ``Note on Varying Speed of Light Cosmologies,''
    Gen. Rel. Grav. \textbf{39} (2007), 511-520
    [arXiv:astro-ph/0703751].

\bibitem{magueijo08}
    J.~Magueijo and J.~W.~Moffat,
    ``Comments on 'Note on varying speed of light theories',''
    Gen. Rel. Grav. \textbf{40} (2008), 1797-1806
    [arXiv:0705.4507].
    
\bibitem{cruz12}
    C.~N.~Cruz and A.~C.~A.~d.~Faria,
    ``Variation of the speed of light with temperature of the expanding universe,''
    Phys. Rev. D \textbf{86} (2012), 027703
    [arXiv:1205.2298].

\bibitem{moffat16}
    J.~W.~Moffat,
    ``Variable Speed of Light Cosmology, Primordial Fluctuations and Gravitational Waves,''
    Eur. Phys. J. C \textbf{76} (2016) no.3, 130
    [arXiv:1404.5567].

\bibitem{franzmann17}
    G.~Franzmann,
    ``Varying fundamental constants: a full covariant approach and cosmological applications,''
    [arXiv:1704.07368].

\bibitem{cruz18}
    C.~N.~Cruz and F.~A.~da Silva,
    ``Variation of the speed of light and a minimum speed in the scenario of an inflationary universe with accelerated expansion,''
    Phys. Dark Univ. \textbf{22} (2018), 127-136
    [arXiv:2009.05397].

\bibitem{costa19}
    R.~Costa, R.~R.~Cuzinatto, E.~M.~G.~Ferreira and G.~Franzmann,
    ``Covariant c-flation: a variational approach,''
    Int. J. Mod. Phys. D \textbf{28} (2019) no.09, 1950119
    [arXiv:1705.03461].

\bibitem{gupta20}
    R.~P.~Gupta,
    ``Cosmology with relativistically varying physical constants,''
    Mon. Not. Roy. Astron. Soc. \textbf{498} (2020) no.3, 4481-4491
    [arXiv:2009.08878].

\bibitem{lee21a}
    S.~Lee,
    ``The minimally extended Varying Speed of Light (meVSL),''
    JCAP \textbf{08} (2021), 054
    [arXiv:2011.09274].
    
\bibitem{lee21b}
    S.~Lee,
    ``Constraints on the time variation of the speed of light using Pantheon dataset,''
    [arXiv:2101.09862].

\bibitem{lee21c}
    S.~Lee,
    ``Constraints on the time variation of the speed of light using Strong lensing,''
    [arXiv:2104.09690].

\bibitem{lee21d}
    S.~Lee,
    ``Cosmic distance duality as a probe of minimally extended varying speed of light,''
    [arXiv:2108.06043].

\bibitem{lee21e}
    S.~Lee,
    ``Determination of varying speed of light from Black hole,''
    [arXiv:2110.08809].

\bibitem{balcerzak14a}
    A.~Balcerzak and M.~P.~Dabrowski,
    ``Redshift drift in varying speed of light cosmology,''
    Phys. Lett. B \textbf{728} (2014), 15-18
    [arXiv:1310.7231].

\bibitem{zhang14}
    P.~Zhang and X.~Meng,
    ``SNe data analysis in variable speed of light cosmologies without cosmological constant,''
    Mod. Phys. Lett. A \textbf{29} (2014), 1450103
    [arXiv:1404.7693].

\bibitem{balcerzak14b}
    A.~Balcerzak and M.~P.~Dabrowski,
    ``A statefinder luminosity distance formula in varying speed of light cosmology,''
    JCAP \textbf{06} (2014), 035
    [arXiv:1406.0150].

\bibitem{qi14}
    J.~Z.~Qi, M.~J.~Zhang and W.~B.~Liu,
    ``Observational constraint on the varying speed of light theory,''
    Phys. Rev. D \textbf{90} (2014) no.6, 063526
    [arXiv:1407.1265].

\bibitem{salzano15}
    V.~Salzano, M.~P.~Dabrowski and R.~Lazkoz,
    ``Measuring the speed of light with Baryon Acoustic Oscillations,''
    Phys. Rev. Lett. \textbf{114} (2015) no.10, 101304
    [arXiv:1412.5653].

\bibitem{salzano16}
    V.~Salzano, M.~P.~D\k{a}browski and R.~Lazkoz,
    ``Probing the constancy of the speed of light with future galaxy survey: The case of SKA and Euclid,''
    Phys. Rev. D \textbf{93} (2016) no.6, 063521
    [arXiv:1511.04732].
    
\bibitem{cai16}
    R.~G.~Cai, Z.~K.~Guo and T.~Yang,
    ``Dodging the cosmic curvature to probe the constancy of the speed of light,''
    JCAP \textbf{08} (2016), 016
    [arXiv:1601.05497].

\bibitem{balcerzak17}
    A.~Balcerzak, M.~P.~Dabrowski and V.~Salzano,
    ``Modelling spatial variations of the speed of light,''
    Annalen Phys. \textbf{529} (2017) no.9, 1600409
    [arXiv:1604.07655].

\bibitem{cao17}
    S.~Cao, M.~Biesiada, J.~Jackson, X.~Zheng, Y.~Zhao and Z.~H.~Zhu,
    ``Measuring the speed of light with ultra-compact radio quasars,''
    JCAP \textbf{02} (2017), 012
    [arXiv:1609.08748].

\bibitem{salzano17a}
    V.~Salzano and M.~P.~Dabrowski,
    ``Statistical hierarchy of varying speed of light cosmologies,''
    Astrophys. J. \textbf{851} (2017) no.2, 97
    [arXiv:1612.06367].

\bibitem{salzano17b}
    V.~Salzano,
    ``How to Reconstruct a Varying Speed of Light Signal from Baryon Acoustic Oscillations Surveys,''
    Universe \textbf{3} (2017) no.2, 35

\bibitem{guedeslang18}
    R.~Guedes Lang, H.~Mart\'\i{}nez-Huerta and V.~de Souza,
    ``Limits on the Lorentz Invariance Violation from UHECR astrophysics,''
    Astrophys. J. \textbf{853} (2018) no.1, 23
    [arXiv:1701.04865].

\bibitem{zou18}
    X.~B.~Zou, H.~K.~Deng, Z.~Y.~Yin and H.~Wei,
    ``Model-Independent Constraints on Lorentz Invariance Violation via the Cosmographic Approach,''
    Phys. Lett. B \textbf{776} (2018), 284-294
    [arXiv:1707.06367].

\bibitem{martinez-huerta18}
    H.~Mart\'\i{}nez-Huerta [HAWC],
    ``Potential constrains on Lorentz invariance violation from the HAWC TeV gamma-rays,''
    PoS \textbf{ICRC2017} (2018), 868
    [arXiv:1708.03384].

\bibitem{cao18}
    S.~Cao, J.~Qi, M.~Biesiada, X.~Zheng, T.~Xu and Z.~H.~Zhu,
    ``Testing the Speed of Light over Cosmological Distances: The Combination of Strongly Lensed and Unlensed Type Ia Supernovae,''
    Astrophys. J. \textbf{867} (2018) no.1, 50
    [arXiv:1810.01287].

\bibitem{guedeslang19}
    R.~G.~Lang, H.~Mart\'\i{}nez-Huerta and V.~de Souza,
    ``Improved limits on Lorentz invariance violation from astrophysical gamma-ray sources,''
    Phys. Rev. D \textbf{99} (2019) no.4, 043015
    [arXiv:1810.13215].

\bibitem{wang19}
    D.~Wang, H.~Zhang, J.~Zheng, Y.~Wang and G.~B.~Zhao,
    ``A model independent constraint on the temporal evolution of the speed of light,''
    [arXiv:1904.04041].

\bibitem{HAWC20}
    A.~Albert \textit{et al.} [HAWC],
    ``Constraints on Lorentz Invariance Violation from HAWC Observations of Gamma Rays above 100 TeV,''
    Phys. Rev. Lett. \textbf{124} (2020) no.13, 131101
    [arXiv:1911.08070].

\bibitem{pan20}
    Y.~Pan, J.~Qi, S.~Cao, T.~Liu, Y.~Liu, S.~Geng, Y.~Lian and Z.~H.~Zhu,
    ``Model-independent constraints on Lorentz invariance violation: implication from updated Gamma-ray burst observations,''
    Astrophys. J. \textbf{890} (2020), 169
    [arXiv:2001.08451].

\bibitem{nguyen20}
    H.~Nguyen,
    ``Analyzing Pantheon SNeIa data in the context of Barrow's variable speed of light,''
    [arXiv:2010.10292].

\bibitem{gupta21}
    R.~P.~Gupta,
    ``Testing the Speed of Light Variation with Strong Gravitational Lensing of SNe 1a,''
    Res. Notes AAS \textbf{5} (2021) no.7, 176

\bibitem{agrawal21}
    R.~Agrawal, H.~Singirikonda and S.~Desai,
    ``Search for Lorentz Invariance Violation from stacked Gamma-Ray Burst spectral lag data,''
    JCAP \textbf{05} (2021), 029
    [arXiv:2102.11248].

\bibitem{mendonca21}
    I.~E.~C.~R.~Mendon\c{c}a, K.~Bora, R.~F.~L.~Holanda, S.~Desai and S.~H.~Pereira,
    ``A search for the variation of speed of light using galaxy cluster gas mass fraction measurements,''
    JCAP \textbf{11} (2021), 034
    [arXiv:2109.14512].
    
\bibitem{hogg99}
    D.~W.~Hogg,
    ``Distance measures in cosmology,''
    [arXiv:astro-ph/9905116].
    
\bibitem{weinberg72}
    S.~Weinberg,
    ``Gravitation and Cosmology: Principles and Applications of the General Theory of Relativity,''

\bibitem{divalentino19}
    E.~Di Valentino, A.~Melchiorri and J.~Silk,
    ``Planck evidence for a closed Universe and a possible crisis for cosmology,''
    Nature Astron. \textbf{4} (2019) no.2, 196-203
    [arXiv:1911.02087].

\bibitem{handley21}
    W.~Handley,
    ``Curvature tension: evidence for a closed universe,''
    Phys. Rev. D \textbf{103} (2021) no.4, L041301
    [arXiv:1908.09139].
    
\bibitem{magana18}
    J.~Magana, M.~H.~Amante, M.~A.~Garcia-Aspeitia and V.~Motta,
    ``The Cardassian expansion revisited: constraints from updated Hubble parameter measurements and type Ia supernova data,''
    Mon. Not. Roy. Astron. Soc. \textbf{476} (2018) no.1, 1036-1049
    [arXiv:1706.09848].
    
\bibitem{moresco22}
    M.~Moresco, L.~Amati, L.~Amendola, S.~Birrer, J.~P.~Blakeslee, M.~Cantiello, A.~Cimatti, J.~Darling, M.~Della Valle and M.~Fishbach, \textit{et al.}
    ``Unveiling the Universe with Emerging Cosmological Probes,''
    [arXiv:2201.07241].

\bibitem{bassett04}
    B.~A.~Bassett and M.~Kunz,
    ``Cosmic distance-duality as a probe of exotic physics and acceleration,''
    Phys. Rev. D \textbf{69} (2004), 101305
    [arXiv:astro-ph/0312443].

\bibitem{holanda10}
    R.~F.~L.~Holanda, J.~A.~S.~Lima and M.~B.~Ribeiro,
    ``Testing the Distance-Duality Relation with Galaxy Clusters and Type Ia Supernovae,''
    Astrophys. J. Lett. \textbf{722} (2010), L233-L237
    [arXiv:1005.4458].

\bibitem{holanda12}
    R.~F.~L.~Holanda, R.~S.~Gon\c{c}alves and J.~S.~Alcaniz,
    ``A test for cosmic distance duality,''
    JCAP \textbf{06} (2012), 022
    [arXiv:1201.2378].

\bibitem{santos-da-costa15}
    S.~Santos-da-Costa, V.~C.~Busti and R.~F.~L.~Holanda,
    ``Two new tests to the distance duality relation with galaxy clusters,''
    JCAP \textbf{10} (2015), 061
    [arXiv:1506.00145].

\bibitem{holanda16}
    R.~F.~L.~Holanda, V.~C.~Busti and J.~S.~Alcaniz,
    ``Probing the cosmic distance duality with strong gravitational lensing and supernovae Ia data,''
    JCAP \textbf{02} (2016), 054
    [arXiv:1512.02486].

\bibitem{goncalves20}
    R.~S.~Gon\c{c}alves, S.~Landau, J.~S.~Alcaniz and R.~F.~L.~Holanda,
    ``Variation in the fine-structure constant and the distance-duality relation,''
    JCAP \textbf{06} (2020), 036
    [arXiv:1907.02118].

\bibitem{mukherjee21}
    P.~Mukherjee and A.~Mukherjee,
    ``Assessment of the cosmic distance duality relation using Gaussian process,''
    Mon. Not. Roy. Astron. Soc. \textbf{504} (2021) no.3, 3938-3946
    [arXiv:2104.06066].

\bibitem{bora21}
    K.~Bora and S.~Desai,
    ``A test of cosmic distance duality relation using SPT-SZ galaxy clusters, Type Ia supernovae, and cosmic chronometers,''
    JCAP \textbf{06} (2021), 052
    [arXiv:2104.00974].
    
\bibitem{holanda21}
    R.~F.~L.~Holanda, F.~S.~Lima, A.~Rana and D.~Jain,
    ``Strong lensing systems and galaxy cluster observations as probe to the cosmic distance duality relation,''
    Eur. Phys. J. C \textbf{82} (2022) no.2, 115
    [arXiv:2104.01614].
    
\bibitem{seikel12} 
  M.~Seikel, C.~Clarkson and M.~Smith,
   ``Reconstruction of dark energy and expansion dynamics using Gaussian processes,''
  JCAP {\bf 1206}, 036 (2012)
  [arXiv:1204.2832].\\
  GaPP is available at \url{https://github.com/astrobengaly/GaPP}
  
\bibitem{shafieloo12} 
    A.~Shafieloo, A.~G.~Kim and E.~V.~Linder,
    ``Gaussian Process Cosmography,''
    Phys.\ Rev.\ D {\bf 85}, 123530 (2012)
    [arXiv:1204.2272].
    
\bibitem{yahya13}
  S.~Yahya, M.~Seikel, C.~Clarkson, R.~Maartens and M.~Smith,
   ``Null tests of the cosmological constant using supernovae,''
  Phys.\ Rev.\ D {\bf 89},  023503 (2014)
  [arXiv:1308.4099].
  
\bibitem{sapone14} 
  D.~Sapone, E.~Majerotto and S.~Nesseris,
  ``Curvature versus distances: Testing the FLRW cosmology,''
  Phys.\ Rev.\ D {\bf 90}, no. 2, 023012 (2014)
  [arXiv:1402.2236].
  
\bibitem{busti14} 
  V.~C.~Busti, C.~Clarkson and M.~Seikel,
  ``Evidence for a Lower Value for $H_0$ from Cosmic Chronometers Data?,''
  Mon.\ Not.\ Roy.\ Astron.\ Soc.\  {\bf 441}, 11 (2014)
  [arXiv:1402.5429].

\bibitem{gonzalez16} 
  J.~E.~Gonzalez, J.~S.~Alcaniz and J.~C.~Carvalho,
   ``Non-parametric reconstruction of cosmological matter perturbations,''
   JCAP {\bf 1604}, 016 (2016)
   [arXiv:1602.01015].
    
\bibitem{marra18} 
  V.~Marra and D.~Sapone,
  ``Null tests of the standard model using the linear model formalism,''
  Phys.\ Rev.\ D {\bf 97}, no. 8, 083510 (2018)
  [arXiv:1712.09676].
  
\bibitem{gomezvalent18a}
    A.~G\'omez-Valent and L.~Amendola,
    ``$H_0$ from cosmic chronometers and Type Ia supernovae, with Gaussian Processes and the novel Weighted Polynomial Regression method,''
    JCAP \textbf{04} (2018), 051
    [arXiv:1802.01505].

\bibitem{gomezvalent18b} 
  A.~G\'omez-Valent,
  ``Quantifying the evidence for the current speed-up of the Universe with low and intermediate-redshift data. A more model-independent approach,''
  JCAP {\bf 1905}, 026 (2019)
  [arXiv:1810.02278].

\bibitem{benisty21}
  D.~Benisty,
  ``Quantifying the $\sigma_8$ tension with model independent approach,''
  Phys. Dark Univ. \textbf{31} (2021), 100766
  [arXiv:2005.03751].
  
\bibitem{bengaly21a}
    C.~A.~P.~Bengaly, C.~Clarkson, M.~Kunz and R.~Maartens,
    ``Null tests of the concordance model in the era of Euclid and the SKA,''
    Phys. Dark Univ. \textbf{33} (2021), 100856
    [arXiv:2007.04879].
  
\bibitem{mukherjee20a}
  P.~Mukherjee and N.~Banerjee,
 ``Non-parametric reconstruction of the cosmological \textit{jerk} parameter,''
  Eur. Phys. J. C \textbf{81} (2021) no.1, 36
  [arXiv:2007.10124].
  
\bibitem{mukherjee20b}
  P.~Mukherjee and N.~Banerjee,
 ``Revisiting a non-parametric reconstruction of the deceleration parameter from observational data,''
  [arXiv:2007.15941].
 
\bibitem{bonilla21}
    A.~Bonilla, S.~Kumar and R.~C.~Nunes,
    ``Measurements of $H_0$ and reconstruction of the dark energy properties from a model-independent joint analysis,''
    Eur. Phys. J. C \textbf{81} (2021) no.2, 127
    [arXiv:2011.07140].

\bibitem{vonMarttens21}
    R.~von Marttens, J.~E.~Gonzalez, J.~Alcaniz, V.~Marra and L.~Casarini,
    ``A model-independent reconstruction of dark sector interactions,''
    Phys. Rev. D \textbf{104} (2021) no.4, 043515
    [arXiv:2011.10846].

\bibitem{colgain21}
    E.~Colg\'ain and M.~M.~Sheikh-Jabbari,
    ``Elucidating cosmological model dependence with $H_0$,''
    Eur. Phys. J. C \textbf{81} (2021) no.10, 892
    [arXiv:2101.08565].
    
\bibitem{arjona21}
    R.~Arjona and S.~Nesseris,
    ``Novel null tests for the spatial curvature and homogeneity of the Universe and their machine learning reconstructions,''
    Phys. Rev. D \textbf{103} (2021) no.10, 103539
    [arXiv:2103.06789].
    
\bibitem{perenon21}
    L.~Perenon, M.~Martinelli, S.~Ili\'c, R.~Maartens, M.~Lochner and C.~Clarkson,
   ``Multi-tasking the growth of cosmological structures,''
    Phys. Dark Univ. \textbf{34} (2021), 100898
    [arXiv:2105.01613].
    
\bibitem{bernardo21a}
    R.~C.~Bernardo and J.~Levi Said,
    ``A data-driven Reconstruction of Horndeski gravity via the Gaussian processes,''
    JCAP \textbf{09} (2021), 014
    [arXiv:2105.12970].
    
\bibitem{bernardo21b}
    R.~C.~Bernardo and J.~Levi Said,
    ``Towards a model-independent reconstruction approach for late-time Hubble data,''
    JCAP \textbf{08} (2021), 027
    [arXiv:2106.08688].
    
\bibitem{bengaly21b}
    C.~Bengaly,
    ``A null test of the Cosmological Principle with BAO measurements,''
    Phys. Dark Univ. \textbf{35} (2022), 100966
    [arXiv:2111.06869].

\bibitem{bernardo21c}
    R.~C.~Bernardo, D.~Grand\'on, J.~L.~Said and V.~H.~C\'ardenas,
    ``Parametric and nonparametric methods hint dark energy evolution,''
    [arXiv:2111.08289].
    
\bibitem{dialektopoulos21}
    K.~Dialektopoulos, J.~L.~Said, J.~Mifsud, J.~Sultana and K.~Z.~Adami,
    ``Neural Network Reconstruction of Late-Time Cosmology and Null Tests,''
    JCAP \textbf{02} (2022) no.02, 023
    [arXiv:2111.11462].
    
\bibitem{avila22}
    F.~Avila, A.~Bernui, A.~Bonilla and R.~C.~Nunes,
    ``Inferring $S_8(z)$ and $\gamma(z)$ with cosmic growth rate measurements using machine learning,''
    [arXiv:2201.07829].

\bibitem{benitez14}
    N.~Benitez \textit{et al.} [J-PAS],
    ``J-PAS: The Javalambre-Physics of the Accelerated Universe Astrophysical Survey,''
    [arXiv:1403.5237].

\bibitem{aparicioresco20}
    M.~Aparicio Resco, A.~L.~Maroto, J.~S.~Alcaniz, L.~R.~Abramo, C.~Hern\'andez-Monteagudo, N.~Ben\'\i{}tez, S.~Carneiro, A.~J.~Cenarro, D.~Crist\'obal-Hornillos and R.~A.~Dupke, \textit{et al.}
    ``J-PAS: forecasts on dark energy and modified gravity theories,''
    Mon. Not. Roy. Astron. Soc. \textbf{493} (2020) no.3, 3616-3631
    [arXiv:1910.02694].

\bibitem{peebles21}
    P.~J.~E.~Peebles,
    ``Improving Physical Cosmology: An Empiricist's Assessment,''
    [arXiv:2106.02672].    


\end{thebibliography}

\end{document}